\documentclass{svproc}
\usepackage{url}
\usepackage{subcaption}
\usepackage{adjustbox}
\usepackage{multirow}
\usepackage{amsmath}
\usepackage{amsfonts}

\let\subparagraph\paragraph
\usepackage[compact]{titlesec}         
\titlespacing{\section}{1pt}{*0}{*0}
\titlespacing{\subsection}{1pt}{*0}{*0}
\titlespacing{\subsubsection}{1pt}{*0}{*0}

\begin{document}
\mainmatter              
\title{Inferring Parsimonious Coupling Statistics in Nonlinear Dynamics with Variational Gaussian Processes}
\titlerunning{Nonlinear Coupling and Variational Approximations}  
%
\author{Ameer Ghouse\inst{1} \and Gaetano Valenza\inst{1}}
\authorrunning{Ameer Ghouse et al.} 
%
\tocauthor{Ameer Ghouse, Gaetano Valenza}
\institute{Department of Information Engineering \& Bioengineering and Robotics Research Center E. Piaggio, School of Engineering, University of Pisa, Pisa, Italy,\\
\email{a.ghouse@studenti.unipi.it}}

\maketitle              

\begin{abstract}
Falsification is the basis for testing existing hypotheses, and a great danger
is posed when results incorrectly reject our prior notions (false positives). Though nonparametric and nonlinear
exploratory methods of uncovering coupling provide a flexible framework to
study network configurations and discover causal graphs,
multiple comparisons analyses make false
positives more likely, exacerbating the need for their control. We aim to
robustify the Gaussian Processes Convergent Cross-Mapping (GP-CCM) method 
through Variational Bayesian Gaussian Process modeling (VGP-CCM). We alleviate
computational costs of integrating with conditional hyperparameter distributions
through mean field approximations. This approximation model,
in conjunction with permutation sampling of the null
distribution, permits significance statistics that are
more robust than permutation sampling with point
hyperparameters. Simulated unidirectional 
Lorenz-Rossler systems as well as mechanistic models of neurovascular systems
are used to evaluate the method. The results demonstrate that the proposed method yields
improved specificity, showing promise to combat false positives.
\dots
\keywords{Dynamical Systems, Causal Analysis, Robust Methods}
\end{abstract}

\section{Introduction}

Coupling measures derived from data allow discovery of system
characteristics and information flow when the experiment does not permit intervention
in the physical process. Traditionally, pairwise coupling metric such as mutual
information or correlation are used for assessing coupling between
systems \cite{Stankovski2017}, such
as for assessing functional connectivity in
the brain \cite{Friston2011}. Dynamical system methods have been popular
for uncovering coupling, such as Granger causality and its nonparametric
extension, transfer entropy \cite{Granger1969,Schreiber2000,Barnett2009}. However, in their standard form, they assess linear
relationships through multivariate autoregression models \cite{Schreiber2000},
though extensions have been made for assessing nonlinear coupling through kernel
methods \cite{Marinazzo2008}. Nonetheless, such methods have been shown to
report spurious detections of causality, i.e. false positives
\cite{Smirnov2012,Smirnov2013}. Straining out the possibilities of false
positive is imperative in hypothesis driven analysis to avoid improper
conclusions based on a priori expectations \cite{Popper1963}--in causal
analysis, the prior notion is non causal relationship between two variables.

Convergent cross-mapping was introduced to account for cases when synergistic
nonlinear coupling may emerge from complex dynamical
systems \cite{Sugihara2012}. This phenomenon is possible to be exploited as,
holistically, the history of state space is assessed (nonparametrically) rather than being
purely an analysis of the lag values' predictive power as with Granger causality.
Formally, convergent cross-mapping (CCM) suggests that, in deterministic systems, there should be
high correlation in predicting a cross-mapped system by taking a state from a state
space reconstruction (as in \cite{Takens1981}) of a putative variable to the
state space reconstruction of a caused variable \cite{Sugihara2012}. Subsequent ameliorations
introduced Gaussian processes with the notion that the
reconstructed state space has uncertainty, thus a prior Gaussian
process mean and covariance functions were applied, either as in \cite{Feng2020} to optimize the
free parameters for state space construction, in \cite{Feng2019}
to analyze residuals in cross-mapping, or to
assess a probability ratio as in \cite{Ghouse2021}. We shall refer to
the latter method in our paper.

Gaussian process models are characterized by mean and covariance (often referred
        to as the kernel) functions \cite{Rasmussen2005}. Though they provide nonparametric inference models for a posteriori
analysis of random processes conditioned on observed data, they are sensitive to the
hyperparameters of the kernel function that describes the covariance between observations. Point
estimates of hyperparameters are optimized by maximum
marginal likelihood of the observations. However, these point estimates may
provide degenerate distributions when generating null distributions derived from
permutation sampling for hypothesis testing as empirically seen in \cite{Ghouse2021} and other contexts of
Bayesian modeling\cite{Chung2015}. We thus propose
placing prior distributions on the hyperparameters and subsequently obtain an approximate
posterior distribution conditioned on observations (who a priori are
zero mean Gaussian processes) using variational Bayesian methods, hereafter
referred to as Variational Gaussian Process Convergent Cross-Mapping (VGP-CCM).
This method avoids expensive Markov Chain Monte Carlo integration
methods for conditional distributions \cite{Blei2017}. With the approximate posterior we integrate out hyperparameters
effects from the coupling statistics and derive nondegenrate null distributions for hypothesis testing. We test
this amelioration on unidirectional coupled Lorenz-Rossler systems and
neurovascular systems.

\section{Materials and Methods}

\subsection{Cross Mapping}
A central tenet of nonlinear analysis in state space using a single time series, which we refer to in lower case letters such as $x$,
hinges on state space reconstruction, where a time series of a variable in a system is viewed as a projection of system's
topology to a single dimension (focusing on reals, $\mathbb{R}^{N\times 1}$ where
N is number of time points), and where using delay-coordinate embedding maps
($\phi(x_i)= \{x_i, x_{i-\tau},x_{i-2\tau},...,x_{i-m\tau}\}$ for i in $1..N-\tau  m$,
where $m$ is an "embedding dimension" and $\tau$ delay time)  of said time
series can reconstruct a state space with isomorphisms to the generating system's structure in
terms of differential geometry \cite{Takens1981,Sauer1991}. Cross-mapping
exploits this theoretical feature of reconstructed state spaces and the nature
of time series being low dimension projections of systems, such that a
state space reconstructed by a putative variable ($\phi(x)$) should be independent of a
caused variable ($\phi(y)$, implying its reconstructed state space contains no dynamical information of the caused
variable leading to the independence of predictions of a caused variable by a putative variable's
state. On the other hand, a prediction of a putative
variable from a caused variable's state space should not be independent of the
putative variable as its reconstructed geometry needs to have topology
consistent with that of the putative variable to predict itself. Convergent
cross-mapping in its original form approaches this problem as a regression task, particularly as
performed by simplex regression \cite{Sugihara2012,Sugihara1990Simplex}.

\subsection{Gaussian Process Convergent Cross Mapping}

Introduced in \cite{Ghouse2021}, GP-CCM utilizes probability ratio to determine the most likely coupling direction based on
spatial analysis in state space through incorporating a delay-coordinate map
into the kernel function that defines the a priori covariance function of the
Gaussian process, i.e. $Cov(x_i,x_j) = K(\phi(x_i),\phi(x_j))$. In other words,
rather than assess predictive powers through regression analysis, GP-CCM
proposed a
probabilistic approach that relates the cross-mapping power as a multivariate probability
distribution. Let $X$ and $Y$ be stochastic processes that
generate the observed time series of $x$ and $y$ respectively. Probability functions
for $X$ and $Y$ are derived by posterior analysis from conditioning a joint Gaussian
Processes by the observed time series $y$ and $x$ respectively.
Consider $\tilde X$ and $\tilde Y$ the null distributions for uncoupled
processes (e.g. through permutation sampling). Then, the null GP-CCM probability ratio
test would be:

\begin{equation}
\label{eq:kappa}
\tilde \kappa_k(X,Y) = \frac{P(\tilde Y|X;\theta_x)}{P(\tilde X|Y;\theta_y)}
\end{equation}

$\theta_x$ or $\theta_y$ are considered nonrandom point
values, $\tilde \kappa$ is a function of the null distribution,
and $k$ is the permutation sample iteration. For sake of conciseness, we
shall abuse the notation, e.g. $P(X|Y)$, for a conditioned
distribution, e.g. $P(X|Y=y)$, for the rest of this paper.

\paragraph{Information Theoretic Results}

In  \cite{Ghouse2021}, the maximum a posteriori (MAP) of the distributions $P(\tilde Y|X;\theta_x)$
and $P(\tilde X|Y;\theta_y)$ were used to provide scalar results. Given the form of
a normal distribution with dimensionality $d$, mean $\mu$ and covariance $\Sigma$:

\begin{equation}
f_x(X) = ((2\pi)^d|\Sigma|)^{-\frac{1}{2}}\mathrm{exp}((X -
            \mu)\Sigma^{-1}(X-\mu)^\top)
\end{equation}

The MAP would be where $X = \mu$, making
the exponential an identity matrix, thus:

\begin{equation}
MAP(\mathcal{N}(\mu,\Sigma)) =
        ((2\pi)^d|\Sigma|)^{-\frac{1}{2}}
\end{equation}

Which we can see, from taking the negative logarithm, is equivalent to the differential
entropy of a Gaussian distribution with an offset of $\frac{d}{2}$:

\begin{equation}
\begin{aligned}
H(\mathcal{N}(\mu,\Sigma)) &= \frac{d}{2} + \frac{d}{2}log(2\pi) +
\frac{1}{2}log(|\Sigma|)\\
&= -log(MAP(\mathcal{N}(\mu,\Sigma))) + \frac{d}{2}
\end{aligned}
\end{equation}
Let $K_k(\tilde X, \tilde Y)=log(\frac{MAP(P(\tilde Y|X;\theta_x)}{MAP(P(\tilde X)|Y;\theta_y)})$, we then obtain:
\begin{equation}
\label{eq:entropyder}
K_k(\tilde X, \tilde Y) = -H(\tilde Y|X;\theta_x) + H(\tilde X|Y;\theta_y) =
log(\frac{|\Sigma_{\tilde X|Y}|}{|\Sigma_{\tilde Y|X}|})
\end{equation}

An interpretation of our probability ratio test in information theory can
be the difference of the amount of information $X$ provides given we were
informed of $Y$ a priori vs how much information $Y$ provides given we were informed of $X$
a priori. Caused variables should provide minimal information to the causing
variable's state space, thus $H(X|Y)<H(Y|X)$ if $X$ drives Y.

\paragraph{Standardizing Results and Null Distributions}
N samples of $K_k(X,Y)$ can be generated from $N$ permutations. The values $K_k(X,Y)$ are normalized by the number of
samples in the time series and then passed to a hyperbolic tangent function in
order to standardize the values between (-1,1). From the samples of $K_k(X,Y)$, an empirical cumulative
distribution is derived and a 1 sided test is performed to see whether the
probability of the causal statistic is less than an $\alpha$ in the null distribution. $\alpha$
is often chosen to be 0.05. Concretely, to test whether $X$ GP-CCM causes $Y$,
the following is calculated:

\begin{equation}
p = \frac{\sum_{j=1}^N U(K(X,Y) - K_j(\tilde X, \tilde Y))}{N}
\label{eq:pvalRes}
\end{equation}

Where U is the Heaviside function which is only 1 when the measured
statistic for the actual observations $K(X,Y)$ is greater than the permutation samples
$K(\tilde X$,$\tilde Y)$, otherwise 0.

\subsection{Variational Gaussian Process Convergent Cross Mapping}
\paragraph{Variational Approximation of the Hyperparameter Distributions}

The proposed method, Variational Gaussian Process Convergent Cross-Mapping
(VGP-CCM), proposes to treat hyperparameters as random effects to be integrated out of
the model, via Monte Carlo integration, exploiting a proposed a posteriori
distribution of hyperparameters $Q(\theta_x|X)$ that approximate the true a posteriori
distribution $P(\theta_x|X)$ in order to exploit independence structures that make integration
computational feasible rather than resorting to more expensive methods such as Gibbs
conditional sampling of the true posteriori \cite{geman1984stochastic}.

To elaborate, hyperparameters are considered point
estimates with no uncertainty in eq. \ref{eq:kappa}. However, studies have shown that point estimates may not
be stable in a variational sense and can
correspond to models that overfit \cite{Beal2003}. Calculus of variations provides a method to
discover models that exhibit minimal "action" from perturbations in their
function. We utilize this framework to discover functions of our
hyperparameters from which we can draw samples to integrate out
the effects of the hyperparameters in the model subsequently. We modify
eq. \ref{eq:kappa} accordingly:

\begin{equation}
\label{eq:kappaint}
\begin{aligned}
\tilde \kappa &= \frac{\int P(\tilde Y|\theta_x)P(\theta_x|X)d{\theta_x}}{\int P(\tilde X|\theta_y)P(\theta_y|Y)d{\theta_y}}\\
                &= \frac{P(\tilde Y|X)}{P(\tilde X|Y)}
\end{aligned}
\end{equation}

The hyperparameter posterior distribution tend to be intractable to compute
as they depend on the marginal distributions of $P(X)$ or $P(Y)$ (model
evidence) respectively:

\begin{equation}
\label{eq:marg}
\begin{aligned}
P(\theta_x|X) &= \frac{P(X|\theta_x)P(\theta_x)}{\int P(X|\theta_x)p(\theta_x)d{\theta_x}}
\end{aligned}
\end{equation}

We, instead, consider a simpler model $Q(\theta_x)$ that approximates the posterior $P(\theta_x|X)$ based
on mean field approximations, eliminating the dependency of conditioning samples on $X$ while
minimizing the statistical distance in the Kullback-Leibler sense
\cite{Kullback1951} from the true posterior
distribution which is conditioned on X. In other words, we want to substitute a
simple distribution $Q(\theta_x)$ with independence structures for
$P(\theta_x|X)$ in eq. \ref{eq:kappaint} to permit computationally efficient Monte Carlo integration:

\begin{equation}
\label{eq:kappaintapprox}
\tilde \kappa \approx \frac{\int P(\tilde Y|\theta_x)Q(\theta_x)d{\theta_x}}{\int P(\tilde X|\theta_y)Q(\theta_y)d{\theta_y}}
\end{equation}

The best approximation $Q$ can be derived by maximizing
the lower bound for the evidence (ELBO) provided the approximate posterior via
the KL divergence.

\begin{equation}
\label{eq:KL1}
\begin{aligned}
KL&(Q||P) = E_{\theta_x \sim Q}(log(\frac{Q(\theta_x)}{P(\theta_x|X)})) =
E_{\theta_x \sim Q}(log(\frac{Q(\theta_x)P(X)}{P(X|\theta_x)P(\theta_x)})\\
         &= log(P(X)) + KL(Q(\theta_q)||P(\theta_q)) -E_{\theta_x \sim Q}(P(X|\theta_x))\\
\end{aligned}
\end{equation}

As the KL divergence is strictly greater than 0, its minimization thereby
is also a maximization of the ELBO.
This can be seen by multiplying both sides by -1 and taking the evidence
($log(P(X))$) and moving it to the left hand side:

\begin{equation}
\label{eq:KL2}
\begin{aligned}
ELBO &=  E_{\theta_x \sim Q}(log(P(X|\theta_x))) - KL(Q(\theta_x)||P(\theta_x))\\
log(P(X)) &- KL(Q||P) = ELBO\\
log(P(X)) & \geq ELBO
\end{aligned}
\end{equation}

Thus our objective to obtain the optimal hyperparameter distributions $Q(\theta_x)$
is to maximize the above ELBO. The next question becomes how to choose the form
of our distribution $Q(\theta_x)$.

\paragraph{Approximate Posterior Form}

The form of the posterior distribution is the art of the practitioner based
on their domain knowledge. For the sake of this
paper, we choose its form to follow a mean-field approximation that
factorizes as $Q(\theta) = \prod_{k=1}^KQ(\theta_k)$ for K hyperparameters.
$\theta$ comprises the hyperparameters of an exponential
squared distance kernel with automatic relevance detection $\sigma$ and $l$
\cite{MacKay1996}, and the inducing pseudopoints for a sparse kernel $z$ \cite{Snelson2005}. In order to
maintain strictly positive parameters for the kernel parameters as well as
having a distribution from which we can reparameterize to draw samples to
automatically compute gradients \cite{Kingma2014} and in which we can derive easily the KL-divergence,
we choose log normal distributions as their priors. The distribution for the inducing
points, instead follows a normal distribution. For the KL divergences in the
ELBO, analytic expressions can be derived. It happens that log normal
$log\mathcal{N}(\mu,\sigma)$ and normal distribution
$\mathcal{N}(\mu,\sigma)$\cite{Kingma2014} share the same expression for KL
divergences.

Note that if the prior P(X) were
chosen independent point distributions, i.e. $P(X)=\delta(X-\mu_p)$, the optimal
$Q(X)$ distribution would have a collapsed mode on $\mu_p$, arriving at the original
formulation of GP-CCM in eq. \ref{eq:kappa} in the case that the prior $\mu_p$ were chosen as the
maximum marginal likelihood solution to $P(X;\theta_x)$.

A priori the hyperparameters are ansatz from the data. We
standardize the data to be zero mean unit variance, thus the covariance
amplitude $\sigma$ has parameters $\mu_{\sigma}=\mathrm{exp}(1)$
and $\sigma_{\sigma}=1$. Length factor $l$ of the kernel has $\mu_{l}=\mathrm{exp}(1)$,
$\sigma_l=1$. The inducing points are a priori mean equal to a random selection of a
subset of the observations with $\sigma_z=1$. Calculus of variations could be
used to derive a coordinate ascent method of obtaining the optimal distribution
Q\cite{Lee2021}. We, instead, choose to maximize the ELBO using
stochastic gradient ascent as in \cite{Kingma2014} and
automatic differentiation in PyTorch \cite{Paszke2017} to avoid manually
deriving the update equations. The expectation $E_{\theta_x \sim Q}(P(X|\theta_x))$ is
approximated by 10 draws of $\theta_x$ from $Q(\theta_x)$ \cite{Kingma2014}.
Furthermore, gradient clipping was used to minimize the effect of the stochastic
gradient's variance\cite{Yoon2018}. The final result from VGP-CCM comes from
applying equation eq. \ref{eq:entropyder} to
the distributions in eq. \ref{eq:kappaint}. Code is publicly
available in Python at the author's Github profile.

\subsection{Synthetic Data}
\paragraph{Unidirectionally Coupled Lorenz and Rossler}
We simulate Rossler (Y) and Lorenz (X) systems with nonlinear unidirectionally coupling
defined by either $\epsilon_x$ or $\epsilon_y$ and diffusion dynamics defined by
a 6d Wiener process W.  The joint system has the
following form:

\begin{equation}
\label{eq:systemofEqs}
\begin{aligned}
 dX_0 &= (\sigma (X_1-X_0)+ \epsilon_y X_0(Y_0 - 1))dt )dt +  \sigma_L(t) dW_{X1}\\
 dX_1 &= (X_0 ( \rho - X_2) - X_1)dt +  \sigma_L(t) dW_{X2}\\
 dX_2 &= (X_0 X_1 - \beta X_2)dt +  \sigma_L(t) dW_{X3}\\
 dY_0 &= (-\omega_2 Y_1 - Y_2 +\epsilon_x Y_0(X_0 - 1))dt +  \sigma_R(t) dW_{Y1}\\
 dY_1 &= (\omega_2 Y_0 + aY_1)dt +  \sigma_R(t) dW_{Y2}\\
 dY_2 &= (b + Y_2(Y1 - c))dt +  \sigma_R(t) dW_{Y3}
\end{aligned}
\end{equation}

The coupling values were chosen as
$\epsilon=\{\epsilon_x,\epsilon_y\}=\{(0,0),(2,0),(4,0),(0,0.2),(0,0.5)\}$.
Thirty realizations of the dynamical system was drawn,
where subsequent permutation analysis was performed to obtain a p-value. The
values of the parameters of the system are used in order to induce deterministic
chaos as seen in table. \ref{tab:parms}.
\begin{table}
\centerline{
\begin{tabular}{|c| c c c c c | c| c c c c c c c|}\hline
Lorenz &  $\sigma_L$ &  $dt$&  $\sigma$&  $\beta$& $\rho$&  Rossler &  $\sigma_R$  & $dt$ & $\omega_1$ & $\omega_2$ & a & b & c\\\hline
 & $10^{-5}$& 0.1 & 10& $\frac{8}{3}$& 28 & & 0.1 & 0.1 & 1.015 & 0.918 & 0.15 &
 0.2 & 10\\\hline
\end{tabular}
}
\caption{Parameters of the Lorenz-Rossler systems}
\label{tab:parms}
\vspace{-4mm}
\end{table}

\paragraph{Unidirectionally Coupled Neurovascular Responses}

The previous system, as mentioned earlier, exhibit highly oscillatory behaviors
specified by its chaotic nature where nearby states may exhibit drastically
divergent trajectories. We also simulate signals which may not be as drastically
oscillatory, though may be emergent of nonlinear phenomenon as seen in
neurovascular signals as observed in neuroscientific studies that exploit
hemodynamics like functional magnetic resonance imaging (fMRI) or functional near
infrared spectroscopy (fNIRS) \cite{stephan2007comparing}.
From \cite{stephan2007comparing}, stimuli generating neurovascular signal ($x$) is
approximated by a bilinear state equation
\begin{equation}
\dot x = Ax + \sum_{j=1}^mu_jB^jx + Cu
\end{equation}

Where the $A$ matrix describes the autonomous undriven dynamics of the system
(here treated as a diagonal matrix with values -1), $u$ are stimuli of a
condition (where the element $u_j$ is 0 when $jth$ event is not
occurring, and 1 when the $jth$ event is occurring), where $B^j$ describes the
interconnectivity between voxels of interest (i.e.
$\begin{bmatrix}\rho_{11}&\rho_{12}\\0&\rho_{22}\end{bmatrix}$ for voxel 1
causing voxel 2, and each $\rho_{ij}$ is an i.i.d. random variable $\rho_{ij}
\sim \mathcal{N}(0,1)$). The B matrices do not change in time, rather they are
realized for each simulated time course. $C$ describes which voxels the stimulus condition interacts with (in this paper,
simulated as a diagonal matrix). In our simulations, we have two conditions
$u_1$, $u_2$ and $u_3$. $u_1$ and $u_2$ are events that occur 10 time each over
a period of 1000 seconds; each event lasts 6 seconds
on intervals of 1 minute. $u_1$ invokes voxel 1 to voxel 2 interactions via
a random variable, $\rho^1_{12} \sim \mathcal{N}(0,1)$ while $u_2$ does not
invoke interaction between the voxels, i.e. $\rho^2_{12} = 0$. $u_3$ is a binomial point process with $p=0.3$ that describes
random neural activity that elicit no functional interconnectivity between
voxels 1 and 2, i.e. $\rho^3_{12}=0$. This serves as background noise where the
activities are occurring on top of.

\begin{figure}
\begin{subfigure}{.49\columnwidth}
\includegraphics[width=\columnwidth]{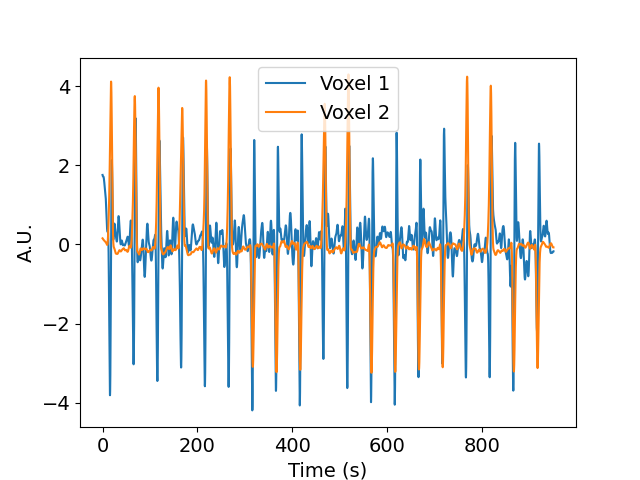}
\caption{Deoxyhemoglobin time courses}
\end{subfigure}
\begin{subfigure}{.49\columnwidth}
\includegraphics[width=\columnwidth]{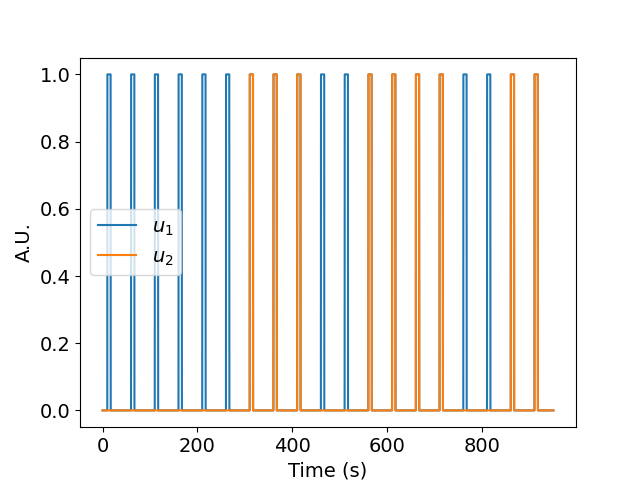}
\caption{Event time courses}
\end{subfigure}
\caption{Exemplary Voxel time course without any SNR observation noise
perturbations}
\label{fig:Voxel}
\end{figure}

After the neural state equations, regional blood flow $f$, blood volume $v$, and
deoxyhemoglobin concentration $q$ is simulated using the proposed parameters and
mechanistic equations in \cite{stephan2007comparing}.
%
%
The nonlinear response of deoxyhemoglobin, $q$, is the neurovascular variable
observed in fMRI and fNIRS. Its emergence from nonlinear equations
provides a complex, nonlinearly saturating signal whose causality may be affected by
parameters of regional properties that make traditional
causal methods hard to apply and begs dynamical systems characterizations
\cite{friston2011dynamic}, and whose characterizations is vital in connectivity
analysis in psychophysiological data. We simulate q with observational
white noise added to corrupt the SNR to 5dB. 
Fig. \ref{fig:Voxel} demonstrates
an exemplary simulation of these interactions.

\subsection{Statistical Analysis}

For VGP-CCM, we generate the null distribution of the coupling statistic using
equation 2, substituting $P(\theta_x|X)$ and $P(\theta_y|Y)$ with our approximate
distributions $Q(\theta_x)$ and $Q(\theta_y)$ respectively. $\tilde Y$ and
$\tilde X$ are sampled from 30 permutations of the original time series $X$ and Y
respectively. GP-CCM, on the other hand, has its null distribution generated
from eq. \ref{eq:kappa} with its hyperparameters obtained as the maximum a posteriori
point estimate from Q obtained in equation 5. An $\alpha=0.05$ is used as the
threshold for significance in all tests when applied to eq. \ref{eq:pvalRes}. Given the Rossler and Lorenz equations
each have 3 variables, we perform 9 comparisons ($X_i\rightarrow Y_1, X_i\rightarrow Y_2, X_i\rightarrow Y_3$ for i in 1...3).
Thus, we have 270 tests for each coupling value. We then
calculate the specificity, a measure of proneness to type I errors, as
$\frac{\text{N correctly accepted $H_0$}}{\text{N  correctly accepted   $H_0$ + N
    incorrectly rejected $H_0$}}$ \cite{Gaddis1990}. This same procedure is
    applied to the neurovascular system, except 100 time series are realized,
    thus 100 tests for coupling results are performed.

\section{Results}

\begin{table}
\centering{
\begin{tabular}{c | cc | cc }
& \multicolumn{2}{c}{GP-CCM}&\multicolumn{2}{c}{VGP-CCM}\\\hline
Chaotic System&L$\rightarrow$R&R$\rightarrow$L&L$\rightarrow$R&R$\rightarrow$L\\\hline
$\epsilon=(0.00,0.00)$& 172& 94& 0 & 3 \\
$\epsilon=(0.00,0.20)$& 203& 62& 0 & 62 \\
$\epsilon=(0.00,0.50)$& 22& 248& 1 & 158\\
$\epsilon=(2.00,0.00)$& 270 &0 & 207& 0 \\
$\epsilon=(4.00,0.00)$& 270& 0& 210& 0 \\\hline
Neurovascular System& $V_1\rightarrow V_2$ & $V_2\rightarrow V_1$& $V_1\rightarrow V_2$& $V_2\rightarrow V_1$ \\\hline
$B = \begin{bmatrix}\rho_{11} & \rho_{12}\\0&\rho_{22}\end{bmatrix}$& 75& 25& 60& 6
\end{tabular}}
\caption{Table displaying the number of rejected $H_o$ for each coupling values
    and each directionality test for the chaotic systems (Rossler to Lorenz equations
        (R$\rightarrow$L),Lorenz to Rossler equations (L$\rightarrow$ R)) or the
        neurovascular system (Voxel 1 to Voxel 2 ($V_1 \rightarrow V_2$) or Voxel 2 to Voxel 1 ($V_2 \rightarrow V_1$). First 2 columns are results obtained from
    results either using point estimates of the hyperparameters (GP-CCM),
    while the last two columns are results using approximate posterior distributions over hyperparameters (VGP-CCM).}
\label{tab:sigvals}
\vspace{-8mm}
\end{table}

Table \ref{tab:sigvals} displays the set of significance rates for various couplings values
for the unidirectionally coupled Lorenz-Rossler systems and for the
neurovascular system. For zero coupling,
VGP-CCM has only rejects the null hypothesis in 3 out of
270 realizations, whereas GP-CCM
rejects the null hypothesis all but 4 of its test. Furthermore, VGP-CCM is more conservative
than GP-CCM in dictating whether to make a claim of significant coupling at low
coupling values, thus VGP-CCM very rarely made a false directionality
claim. For the case of detecting Rossler driving Lorenz
coupling, VGP-CCM reports a specificity of 99.6\%
while GP-CCM reports a specificity
of 58.3\%. In the direction from Lorenz driving Rossler,
GP-CCM and VGP-CCM both report a specificity of 100\%. For correctly rejecting the hypothesis in
either direction, VGP-CCM has a specificity of 99.8\%
compared to GP-CCM's 79.1\%.  Similarly for the neurovascular systems, the
proposed variational VGP-CCM provides more specificity at 94\% compared to
GP-CCM's 75\%.

\begin{figure}
\begin{subfigure}{.5\columnwidth}
\includegraphics[width=\linewidth]{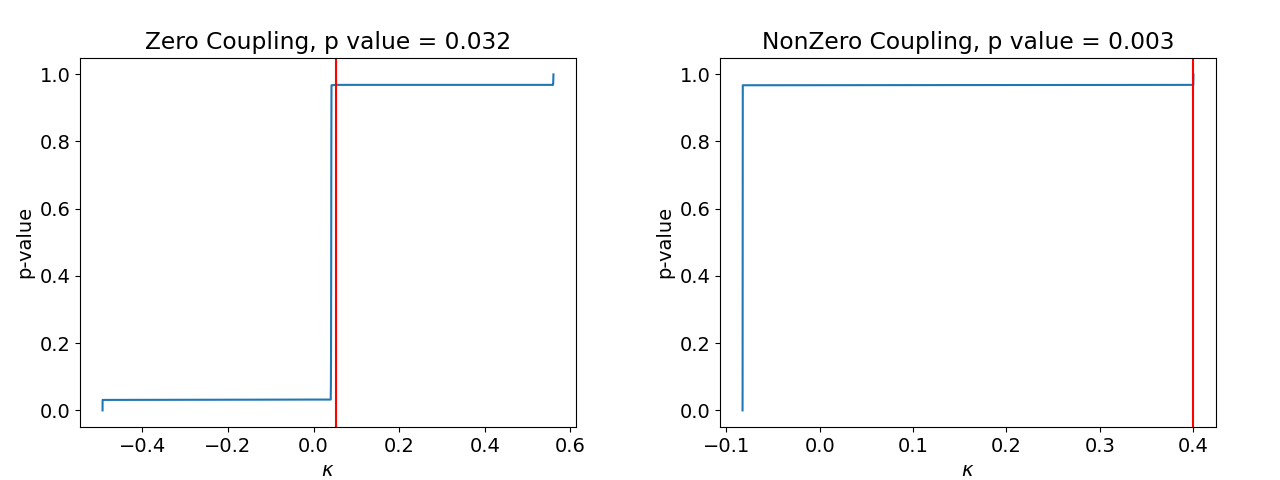}
\caption{Original GP-CCM method}
\end{subfigure}
\begin{subfigure}{.5\columnwidth}
\includegraphics[width=\linewidth]{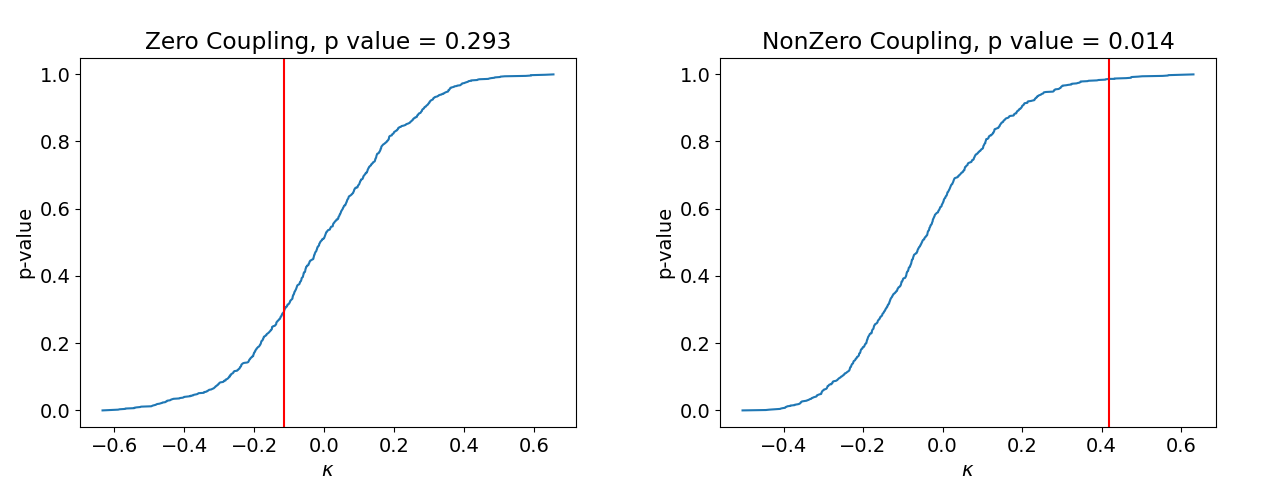}
\caption{Variational GP-CCM method}
\end{subfigure}
\caption{Example null distributions and p values for zero and non zero coupling
    seen from the Lorenz system to the Rossler system using the original GP-CCM method (a) and the proposed variational GP-CCM
        method (b). The value for nonpermuted maximum a posteriori time series results is shown in red}
\label{fig:cumdist}
\end{figure}

Fig. \ref{fig:cumdist} displays cumulative distributions for simulating a Lorenz
system of equations driving Rossler equations for coupling values
$\epsilon=\{(0,0),(2,0)\}$ between variables X0 and Y0. The cumulative distribution reveals a robust null distribution when
using the approximate posterior compared to using fixed values for the
hyperparameters and performing permutations where results from only permuting
samples results in a degenerate distribution with a collapsed mode. Both GP-CCM
and VGP-CCM have an increasing raw coupling statistic as the coupling value
increases.

\section{Discussion}

We propose the VGP-CCM framework, which extends the GP-CCM to use a
variational approximation to the posterior distribution of hyperparameters.  The
presented study aimed to use a variational approximation to the posterior distribution of
hyperparameters to develop a more parsimonious distribution for testing a null
uncoupled hypothesis for GP-CCM. Tests were performed on simulated unidirectionally
coupled Lorenz-Rossler systems to control the values of coupling between the
variables of the system of equations to determine whether GP-CCM and VGP-CCM
suffered from false positives in the significance tests and at what coupling
could either make significance statements. Neurovascular systems were also
simulated to illustrate VGP-CCM's efficacy at uncovering coupling in more slowly
oscillating nonlinear signals compared to the chaotic maps studied in the Lorenz-Rossler
system.

From the results over 30 realizations, as seen in table \ref{tab:sigvals},
VGP-CCM has much improved results compared to GP-CCM, with nearly
20\% greater specificity than GP-CCM. Furthermore, from the simulations of the neurovascular system, we demonstrated
that the proposed variational extension continues to provide further robustness
even for slow oscillating signals as observed in deoxyhemoglobin signals, again
with a similar substantial increase in specificity. As mentioned in the introduction, indeed
the degenerate nature of the null distribution, as seen in fig
\ref{fig:cumdist}a, allowed too much liberty to make
claims of significance even in the wrong direction. This proves promising in the
case that even in low coupling, VGP-CCM will not make a claim to the wrong
direction, giving researchers confidence in the results that if the raw
statistic is insignificant but with high value, it may be a matter of needing to collect
more realizations, as seen in the low couplings values in table
\ref{tab:sigvals}.

The prior distribution used in this study were indeed uninformative naively,
though they allowed ease of computations, reparameterizations of means
and variations with unit Gaussian random sample draws for automatic
differentiation, and provided domain constraints.  We suggest in studies where
domain knowledge of how samples should a priori correlate is available, such as a kernel derived
from mechanistic dynamic equations--e.g. time constants in linear ordinary
differential equations as performed in dynamic causal modeling
\cite{Friston2003} or temperature in thermodynamic heat
systems\cite{Berline1996}--could be used. Otherwise, a separate dataset with
labels can be used to infer optimal distributions through cross-validations.
Even further directions may consider generative models
for the posterior distribution conditioned on data, such as neural
networks\cite{MacKay1996,Blei2017}.

From this study, we can conclude that integrating out the hyperparameters from
the GP-CCM probability ratio may result in a more parsimonious statistic giving
robust null distributions for hypothesis testings. The computational aspects are
lessened by avoiding Markov Chain Monte Carlo integration on the true
conditional a posteriori distribution, instead opting for variational approximations of the
hyperparameter distribution with independent mean field structures.  The robustness was verified on simulated data from unidirectional Lorenz-Rossler
systems, where VGP-CCM achieved much higher specificity than GP-CCM. Further
studies will look into applying this methodology on real world data such as
neurophysiological systems.

\section{Acknowledgments}

The research leading to these results has received partial funding
from the European Commission - Horizon 2020 Program under grant agreement n°
813234 of the project “RHUMBO”, and by Italian Ministry of Education and
Research (MIUR) in the framework of the CrossLab project (Departments of
Excellence).

\end{document}